\documentclass[conference]{IEEEtran}
\makeatletter
\def\ps@headings{%
\def\@oddhead{\mbox{}\scriptsize\rightmark \hfil \thepage}%
\def\@evenhead{\scriptsize\thepage \hfil \leftmark\mbox{}}%
\def\@oddfoot{}%
\def\@evenfoot{}}
\makeatother
\usepackage{subfigure, verbatim, soul}
\usepackage{comment,pifont,array,url,amsmath,multirow,latexsym,tabularx}
\usepackage{ifpdf,rotating,colortbl}
\usepackage{graphicx}
\usepackage[boxed]{algorithm}
\usepackage{algorithmic}
\usepackage{xspace}
\usepackage{url}
\usepackage{bbm}
\usepackage{wrapfig}

\newcommand{\eat}[1]{}

\newcommand{\cs}{SwarmServer}
\newcommand{\sscs}{CheatSheet}
\newcommand{\ssaf}{AntFarm}
\newcommand{\ssmm}{Leveler}
\newcommand{\ssaiad}{AIAD}
\newcommand{\maxmin}{MIN\_MAX}
\newcommand{\maxavg}{MIN\_AVG}
\newcommand{\mincost}{MIN\_COST}

\newcommand{\btcap} {BitTorrent}
\newcommand{\eqsplit} {EqualSplit}
\newcommand{\propsplit} {PropSplit}
\newcommand{\maxminheur} {MaxMinHeur}
\newcommand{\Cacd} {Client-assisted content delivery}
\newcommand{\cacd} {client-assisted content delivery}

\newcommand{\mudis}{\tilde{\mu}}
\newcommand{\claim}[1]{{\sc{claim} #1}}

\title{Pros \& Cons of  Model-based Bandwidth Control  for Client-assisted Content Delivery}

\author{
{Abhigyan Sharma\quad\quad Arun Venkataramani}\\
University of Massachusetts Amherst
\and
{Antonio A Rocha}\\
Fluminense Federal University, Brazil
} 

\usepackage{amsmath}    
\usepackage{amssymb}
\usepackage{subfigure}

\begin{document}

\maketitle

\begin{abstract}



A key challenge in \cacd\ is determining how to allocate limited server bandwidth across a large number of files being concurrently served so as to optimize global performance and cost objectives. In this paper, we present a comprehensive experimental evaluation of strategies to control server bandwidth allocation. As part of this effort, we introduce a new {\em model-based} control approach that relies on an accurate yet concise ``cheat sheet'' based on a priori offline measurement to predict swarm performance as a function of the server bandwidth and other swarm parameters. Our evaluation using a prototype system, \cs, instantiating static, dynamic, and model-based controllers shows that static and dynamic controllers can both be suboptimal due to different reasons. In comparison, a model-based
approach consistently outperforms both static and dynamic approaches
provided it has access to detailed measurements in the regime of
interest. Nevertheless, the broad applicability of a model-based
approach may be limited in practice because of the overhead of
developing and maintaining a comprehensive measurement-based model of
swarm performance in each regime of interest.

\end{abstract}
\section{Introduction}

Faced with the challenge of ever-increasing demand for content, content distributors have turned to {\em \cacd} in recent times. A \cacd\ architecture enables content distributors to provide performance in a scalable and cost-effective manner by opportunistically leveraging client resources, especially their uplink bandwidth, to augment their managed infrastructure resources. Although \cacd\ systems have their roots in peer-to-peer file sharing systems \cite{bt, pando}, commercial CDNs such as Akamai, Velocix, and Octoshape \cite{akamai, velocix, octoshape} as well as live streaming services such as PPLive and Sopcast \cite{pplive,sopcast} have warmed up to using them for mainstream enterprise content delivery in recent times.

A key problem in \cacd\ is bandwidth management, i.e., determining how to allocate limited server bandwidth across a large number of files being concurrently served to clients so as to balance the performance and cost objectives of the content distributor. Unlike purely client-server systems or purely peer-to-peer systems, this problem is particular to \cacd\ systems that attempt to combine the predictable performance and ease of management of the former with the scalability and cost-effectiveness of the latter. The sever bandwidth allocated to a {\em swarm}, or a set of clients concurrently downloading the same file, is critical in determining the effectiveness of client-to-client exchanges and by consequence client-perceived performance. Furthermore, 
the appropriate allocation may be counter-intuitive, e.g., a popular file requires {\em less} server bandwidth compared to an unpopular file, all else being equal, in order to ensure similar client-perceived performance.

Our primary contribution is a measurement-driven comparative analysis of several existing and new strategies for allocating server bandwidth in \cacd\ systems. To this end, we classify these bandwidth allocation strategies, or {\em controllers}, into three categories. The first is {\em static}, a class of controllers that use simplistic strategies such as allocating bandwidth uniformly, on a best-effort basis, or proportional to the demand across files \cite{bt}. 
The second is {\em dynamic}, a class of controllers that constantly adjust the allocation in response to fine-grained client-perceived performance so as to optimize the performance or cost objectives of the content distributor \cite{antfarm,vformation}.

In this paper, we present a third, new class of controllers called {\em model-based} controllers that allocate server bandwidth based on a predictive model of client-perceived performance as a function of the server bandwidth and other swarm parameters such as the request arrival rate, file size, and client upload capacities. Unlike dynamic controllers that can be suboptimal due to long convergence delays while searching for an optimal allocation in situ, model-based controllers can jump to the optimal allocation in a single step by solving the underlying optimization problem ``on paper''. 

We have implemented a prototype system, \cs, to facilitate our comparative analysis of controllers. In addition to several simple static and dynamic controllers, \cs\ supports a model-based controller called CheatSheet for three bandwidth allocation objectives: minimizing the average download time, maximum download time, or the server bandwidth consumed so as to achieve a target performance objective. CheatSheet uses extensive a priori measurement to develop an accurate and concise model of performance as a function of the server bandwidth and a number of swarm parameters. To our knowledge, CheatSheet is the first attempt at developing a detailed {\em empirical} model of swarm performance. 

Our extensive experiments with \cs\ in conjunction with BitTorrent swarms running over 350 PlanetLab nodes reveal several insights.  First, simple static controllers are hit-or-miss; while they perform well for some performance objectives and workloads, even outperforming dynamic controllers, they fall severely short on others. The suboptimal performance of static controllers is unsurprising and consistent with previous findings \cite{antfarm} for one our three objectives of interest. Second, model-based control is feasible and promising---CheatSheet consistently outperforms both static and dynamic controllers provided its model is based on detailed a priori measurements in an environment similar to the operational environment. CheatSheet performs up to 4$\times$ better than static schemes and up to 1.7$\times$ better than dynamic controllers.

Nevertheless, having gone through the experience of building a model-based controller, our conclusions about its practicality are somewhat mixed because of several reasons. First, it is hard. To appreciate this, consider that CheatSheet's model used in the experiments in this paper alone required over 12 days of measurement data on PlanetLab  so as to account for a number of parameters such as the server bandwidth, request arrival rate, distribution of client upload capacities, file size, etc. Second, while a measurement-driven model is robust to small variations in the operational environment, significant changes require recalibrating the model. For example, we find that the model developed over PlanetLab is inaccurate when deployed on a public cloud such as Amazon EC2 or a local cluster in our department. Similarly, significant changes in the client population or behavior such as participation in multiple swarms introduce further uncertainties into the model. Thus, model-based control may be appropriate primarily for relatively predictable environments (e.g., distributing TV shows and movies to FIOS \cite{FIOS} customers).

The rest of the paper quantifies these nuanced pros and cons of the three classes of controllers. We begin with a background on \cacd.

\section{Background}
\label{sec:bg}

\begin{figure*}[t]
\hspace{-0.3cm}
\begin{minipage}[b]{0.6\linewidth}
\centering
\vspace{-1.2in}\includegraphics[scale=0.5]{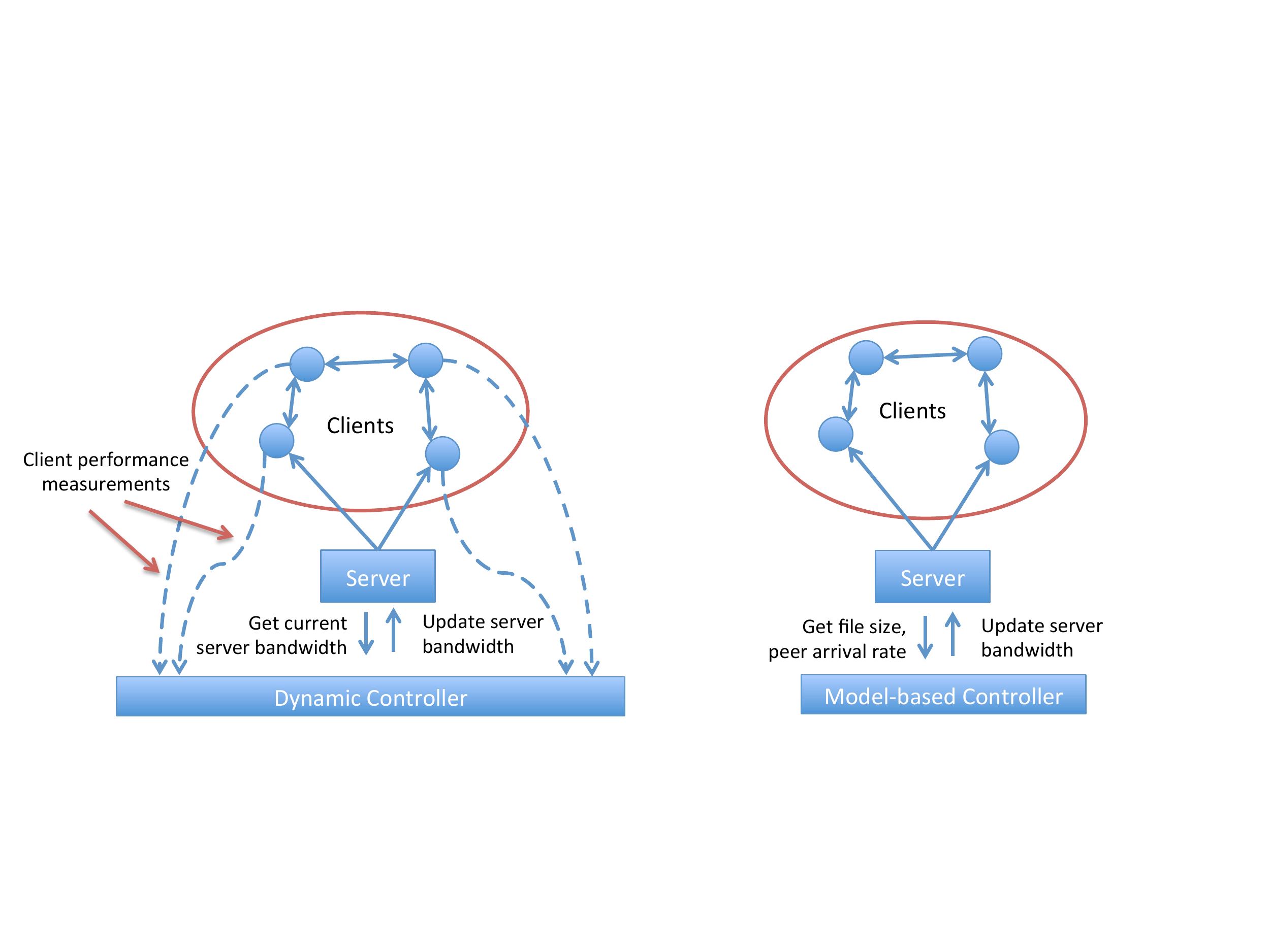}\vspace{-1.38in}
\vspace{0.2in}
\caption{A comparison of dynamic and model-based control architectures.}
\label{fig:dynamicmodel}
\end{minipage}
\begin{minipage}[b]{0.4\linewidth}
\centering
\includegraphics[scale=0.43]{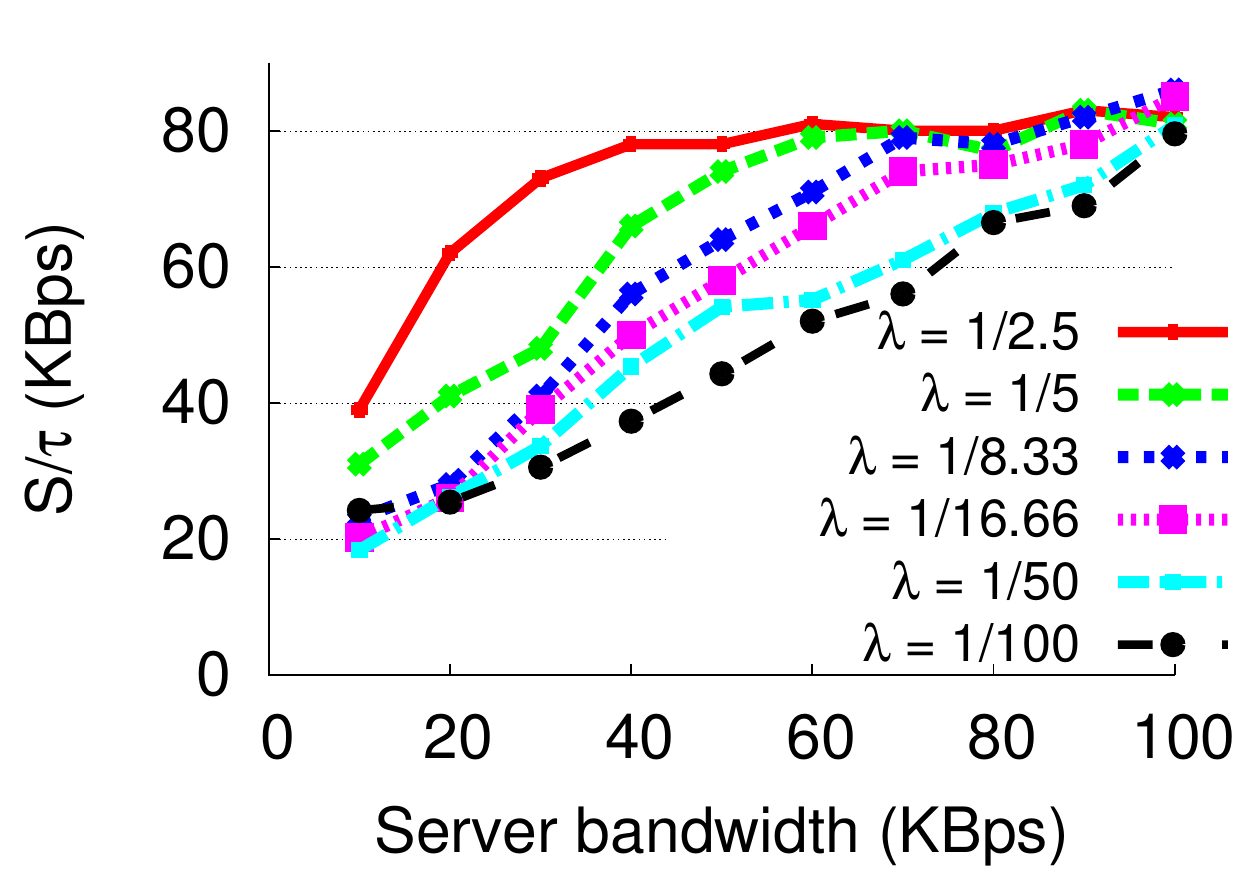}
\caption{Dependence of swarm performance on server bandwidth ($x$), and peer arrival rate $\lambda$ for $S$ = 10MB and $\mu$ = 100KBps.}
\label{fig:basic}\end{minipage}
\vspace{-0.25in}
\end{figure*}

A \cacd\ system consists of a server that acts as the primary source for all content.  All clients concurrently downloading the same file are referred to as a  swarm.  Clients follow a common peer-to-peer protocol for downloading (uploading) the file from (to) other clients in the swarm. The server participates by contributing bandwidth to all swarms. In this paper, we focus on the BitTorrent protocol \cite{bt} because of its open nature and wide deployment, however our findings are qualitatively applicable to other comparable plugins offered by content distributors \cite{akamai,octoshape}.  


A key goal of a \cacd\ system is to optimize a system-wide objective, e.g., minimize the average download time of all clients, by judiciously allocating limited server bandwidth across all swarms. To this end, a {\em controller} at the server collects information from all swarms and uses this information to compute and effect an allocation of server bandwidth so as to optimize the system-wide objective.

\subsection{Classification of controllers}


We classify existing controllers as static or dynamic, and introduce a new class called model-based controllers. 

\textbf{Static:} A static controller allocates server bandwidth using a simple heuristic while being agnostic to the system-wide performance objective and unresponsive to actual client-perceived performance.  For example, a static controller that we analyze is using BitTorrent as-is by repurposing a common seeder across all swarms as the server \cite{bt}.


\textbf{Dynamic:} A dynamic controller continuously monitors fine-grained information about client-perceived performance for all clients in each swarm (see Figure \ref{fig:dynamicmodel}, left), and accordingly adjusts the bandwidth allocation in each monitoring epoch. An example of a dynamic controller is AntFarm \cite{antfarm} that monitors the number of blocks uploaded and downloaded by each client in each  epoch, uses a strategy based on perturbation and gradient-ascent in order to  optimize the aggregate download rate across all clients across all swarms.




\textbf{Model-based:} A model-based controller relies on a predictive model of swarm performance as a function of the supplied server bandwidth and other swarm parameters such as the file size, the peer arrival rate, and the upload capacity distribution of peers. Unlike dynamic controllers, a predictive model obviates explicit measurement of client-perceived performance, requiring only parameters that are already available or easily inferred at the server (see Figure \ref{fig:dynamicmodel}, right). More importantly, it obviates {in situ} perturbation and gradual adjustment of the allocation enabling the controller to jump to the optimal allocation in a single step by using the model to solve the underlying optimization problem ``on paper''. Thus, a model-based controller can quickly adapt to sudden changes in request arrival rates.


\subsection{Limitations of dynamic control}


Our motivation for investigating model-based control stems from the limitations of dynamic controllers in realistic environments. Unlike static controllers that are but naive baseline strategies, the limitations of dynamic control are less obvious and are described next.

\textbf{Convergence time:}  Dynamic control works in a feedback-driven manner by perturbing the current allocation, monitoring the performance impact of the perturbation, and accordingly determining the next perturbation. This approach is prone to prohibitively long convergence delays, primarily because the effect of a perturbed allocation can take several minutes to propagate through the swarm so as to be observable by the controller. As an example, AntFarm updates its server bandwidth once every 300 seconds by 5KB/s, so an adjustment of 50KBps requires nearly an hour to take effect.
 
 \textbf{Measurement overhead:} Dynamic controllers utilize server resources to monitor every client's performance in a swarm; this overhead  can be significant for a swarming system with tens of thousands of clients.

\textbf{Measurement error:} The performance of any controller in steady state depends on how accurately it can estimate  the relation between server bandwidth and swarm performance.
Dynamic controllers can inaccurately estimate this relation because they measure swarm performance for the current bandwidth allocation only for a single measurement interval of a few hundred seconds. The statistical variations in the number of peers joining the swarm and in their upload capacities introduce error in measuring swarm performance.

These limitations of dynamic controllers compel us to explore model-based controllers. We hope that the measurement overhead could be relegated to an a priori offline phase to develop an accurate model of swarm performance 	in exchange for increased responsiveness in the operational phase. The challenge, of course, is to develop an accurate model of swarm performance with a tractable measurement overhead and small representation size, a challenge we address next.

\section{A Measurement-based Model}
\label{sec:model-based}


In this section, we develop a measurement-based model of swarm performance -- the key building block for a model-based controller. Unlike prior theoretical models \cite{qiusrikant,fanchiu,Liao} that over-simplify swarm behavior, our work, to our knowledge, is the first effort at developing a measurement-based model of swarm performance.
Despite our progress, the proposed model falls short both because it requires very extensive measurements lasting several days, but more fundamentally due to the large number of factors that affect swarm performance, even with several simplifying assumptions.





\subsection{Goal and model assumptions}
\label{sec:assumption}

We start with the following question: {\em what is the average download time of peers in a BitTorrent swarm when given a certain amount of server bandwidth}? The answer to this question of course depends on several characteristics of the swarm such as the arrival and departure patterns of peers, their upload and download capacities, the size of the file being distributed, etc. The answer also depends on design parameters of BitTorrent clients such as the number of active peers to which a peer concurrently uploads and how it splits its upload capacity across them, the length of an optimistic unchoke round, the size of chunks, etc. Finally, network conditions and artifacts of the transport protocol (TCP or custom transport protocols such as $\mu$TP for non-interfering downloads \cite{uTP}) will also impact swarm performance.  Clearly, a model attempting to account for all of the factors affecting a swarm's performance quickly becomes intractable. 


To derive a simple yet useful model, we consider a swarm distributing a file of size $S$ to peers arriving at a rate $\lambda$. The  upload capacities of arriving peers are drawn from a distribution with mean $\mu$. The download capacity of peers is unlimited. Peers depart immediately after finishing their download (so the departure rate of peers is equal to the arrival rate $\lambda$ in steady state). Let $x$ denote the (fixed) bandwidth supplied by the server.  Our model postulates that the average download time of peers, $\tau$, can be determined as a function $f$ of $x, \mu, \lambda$ and $S$. We state this dependence as

 \vspace{-0.2in}
\begin{eqnarray}
\frac{S}{\tau} = f(x, \mu, \lambda, S)
\label{eq:model}
\end{eqnarray}

\vspace{-0.1in}

We call $\frac{S}{\tau}$ as \emph{swarm performance}. As the average download time of peers ($\tau$) reduces, swarm performance improves.




By assuming that $\tau$ is determined by the above four parameters alone, the model implicitly makes a few assumptions. The model assumes that network loss rates and round-trip times are not so high that they reduce the effective average peer upload capacity (or equivalently that $\mu$ already incorporates these effects). It also implicitly assumes that all peers use a standard BitTorrent client and that implementation variations across operating systems are minor. It further assumes that $\mu$ already incorporates the effect of user-specific configurations that limit their upload contribution. Finally, the model assumes that despite all these heterogeneous factors affecting the {\em distribution} of peer upload capacities in practice, this distribution is stationary, so the {\em average} upload capacity $\mu$ (in conjunction with the other three parameters) is sufficient to determine the average download time.

\subsection{Measurement-based model}
\label{sec:measurement}


We take an empirical, measurement-driven approach to capture the relationship in Equation (\ref{eq:model}). A naive approach to this end would be to ``measure" the relationship posed in Equation (\ref{eq:model}) for all foreseeable values of the four underlying dimensions ($x, {\mu}, \lambda, S$), which is impractical. Instead, our approach is to summarize the relationship using a small number of measured scenarios and use simple interpolation to estimate the unmeasured scenarios. We begin with a description of our measurement setup.




\subsubsection{Measurement setup}
\label{sec:expsetup}

Our measurement testbed consists of 350 PlanetLab nodes installed with an an instrumented BitTorrent client \cite{legout:07}, and two (non PlanetLab)  servers hosted at our university that act as the seeder and the tracker respectively for all swarms. In each swarm run, peers arrive over time at a PlanetLab node  to download the file and depart immediately after completing the download. Each swarm is run long enough so that the average download times of peers stabilizes, and the server records the average download time of peers that have completed downloads at the end of the experiment.  Each swarm run is repeated five times with a fixed set of parameters $(x, \mu, \lambda, S)$ and different runs vary these parameters.

\eat{
We use PlanetLab for running private swarms in order to obtain measurement data as follows. Two (non PlanetLab) nodes, hosted at our university act as the server and the tracker respectively, while the rest of the nodes (on PlanetLab) act as peers and run an instrumented BitTorrent client \cite{legout:07}. In each swarm run, peers arrive over time to download the file and depart immediately after completing the download. Peer inter-arrival times  follow an exponential distribution with mean $1/\lambda$. Each swarm is run long enough so that the download times of peers stabilize, and the server records the average download time of peers that have completed downloads at the end of the experiment.  Each swarm run is repeated five times with a fixed set of swarm parameters $(x, \mu, \lambda, S)$ and different runs vary these parameters.
}

We use the upload capacity distribution of BitTorrent peers reported in \cite{bittyrant}, which was scaled and truncated to remove very high capacity peers so as to accommodate the daily  limit on the maximum data transfer imposed on PlanetLab nodes. The resulting average upload capacity  ($\mu$) is 100 KBps with upload bandwidths in the range of 40 to 200 KBps for individual peers. No restrictions are imposed on the maximum download rate of any client.  The file size is fixed at $S$ = 10 MB. Peer inter-arrival times are exponentially distributed with mean $1/\lambda$.

Figure \ref{fig:basic} shows the aggregate results of our measurement experiments. Each line corresponds to a fixed arrival rate $\lambda$ as shown, and  plots the swarm performance for different values of the server bandwidth $x$ that is varied from 10 to 100 KBps (also the average peer upload capacity) in 10 KBps increments.  With these parameters, a swarm run takes between 2000 to 5000 seconds, so the total running time to generate this figure is over 12 days (5 runs per point $\times$ 60 points $\times$ an hour roughly per run = 300 hours).

\subsubsection{Swarm performance vs. server bandwidth}



Figure \ref{fig:basic} presents several insights about how the swarm performance depends on server bandwidth and peer arrival rate. First, swarm performance as expected increases with server bandwidth keeping all else fixed. Second, swarm performance is concave with respect to server bandwidth. This is because, when the server bandwidth is very low, it becomes the bottleneck preventing peers from efficiently utilizing their upload capacity for exchanging blocks. In this regime, increasing server bandwidth slightly improves the efficiency of P2P exchanges, which improves swarm performance significantly. At high values of server bandwidth, there is less room for improving the efficiency of P2P exchanges, so the server's bandwidth improves performance similar to traditional client-server systems, i.e., the bandwidth is divided across extant peers. When the server bandwidth equals the average peer upload capacity we find that a swarm's utilization of P2P bandwidth is about as efficient as it can be, and any additional server bandwidth is simply used as in a client-server system. As a result, the swarm performance in the regime $x>\mu$ (not shown in Figure \ref{fig:basic}) can be easily derived analytically obviating time-consuming measurements.




Third, in the regime $x\leq\mu$ shown in the figure, swarm performance improves with the arrival rate (keeping all else fixed). At very low arrival rates, e.g., $\lambda$ = 1/100/s, the swarm behaves like a client-server system as there is at most one peer most of the time, so the corresponding curve resembles the line $y=x$. At higher arrival rates, the swarm remains efficient (i.e., it maintains a healthy download rate of over 80 KBps) for values of $x$ much smaller than $\mu$. This is because large swarms are mostly self-sustaining and need only a tiny amount of server bandwidth to supply missing blocks in the unlikely event that none of the extant peers possess those blocks.


\subsubsection{Model representation}


To concisely represent the swarm-performance model, we carefully select a small number of values of each parameter  for measurements. We maintain a table, referred to as the ``cheat sheet'', that records the swarm performance for all combinations of these parameters.  This cheat sheet is used to approximately estimate by simple linear interpolation the swarm performance for values of parameters that are not explicitly measured. Next, we describe how we select the values of the model parameters for measurements.


\paragraph{Server bandwidth \& peer arrival rate} The dependence of swarm performance on server bandwidth and peer arrival rate for a given upload capacity distribution and file size (as in Figure \ref{fig:basic}) is captured using $\approx$ 100 values.
We take measurements for ten values of $x$ ranging from $\mu/10$ to $\mu$, and for ten values of $\lambda$ in a range determined by a metric we refer to as the ``healthy swarm size''. The healthy swarm size is the number of peers when the efficiency of P2P exchanges in maximum. The intuition for healthy swarm size comes from Little's law \cite{littlelaw}, healthy swarm size is  $\lambda \times S/\mu$, as $S/\mu$ is the average download time of peers in this case. When the healthy swarm size is one or less, the swarm essentially behaves like a client-server system. We empirically observe that when the healthy swarm size is 50 or more, the swarm is essentially self-sustaining, i.e., even with a server bandwidth of just a $\mu/10$, the swarm is efficient. So we take measurements for values of $\lambda$ selected such that the healthy swarm size $\lambda S/\mu$ increases from 1 to 50 in 10 equal increments. The total number of combinations of $x$ and $\lambda$ is therefore 100.




\paragraph{File size}



\vspace{-0.1in}
\begin{figure}[ht]
\begin{minipage}[b]{0.45\linewidth}
\centering
\includegraphics[scale=0.35]{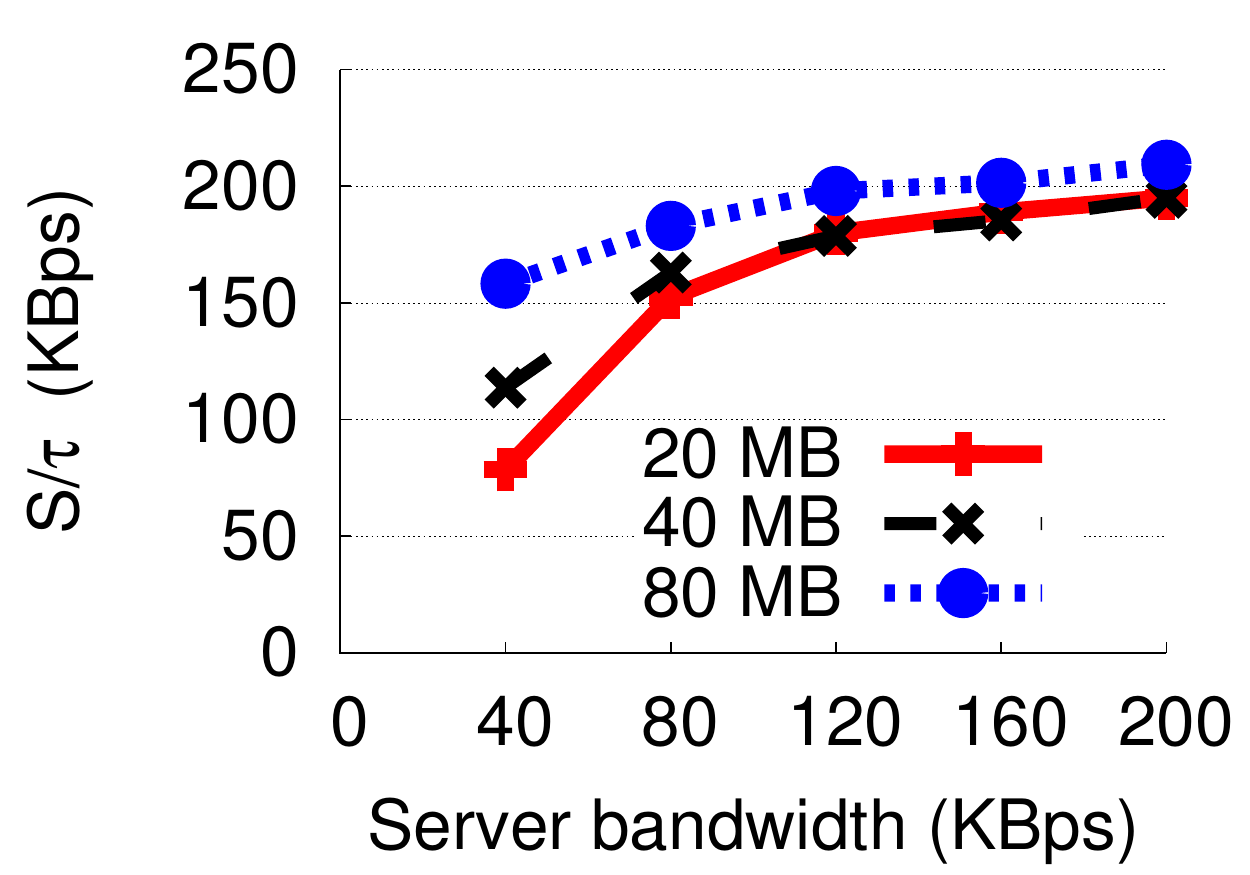}
\caption{Swarm performance improves with file size for same aggregate demand $\lambda S$.}
\label{fig:size}
\end{minipage}
\hspace{0.3cm}
\begin{minipage}[b]{0.45\linewidth}
\centering
\includegraphics[scale=0.35]{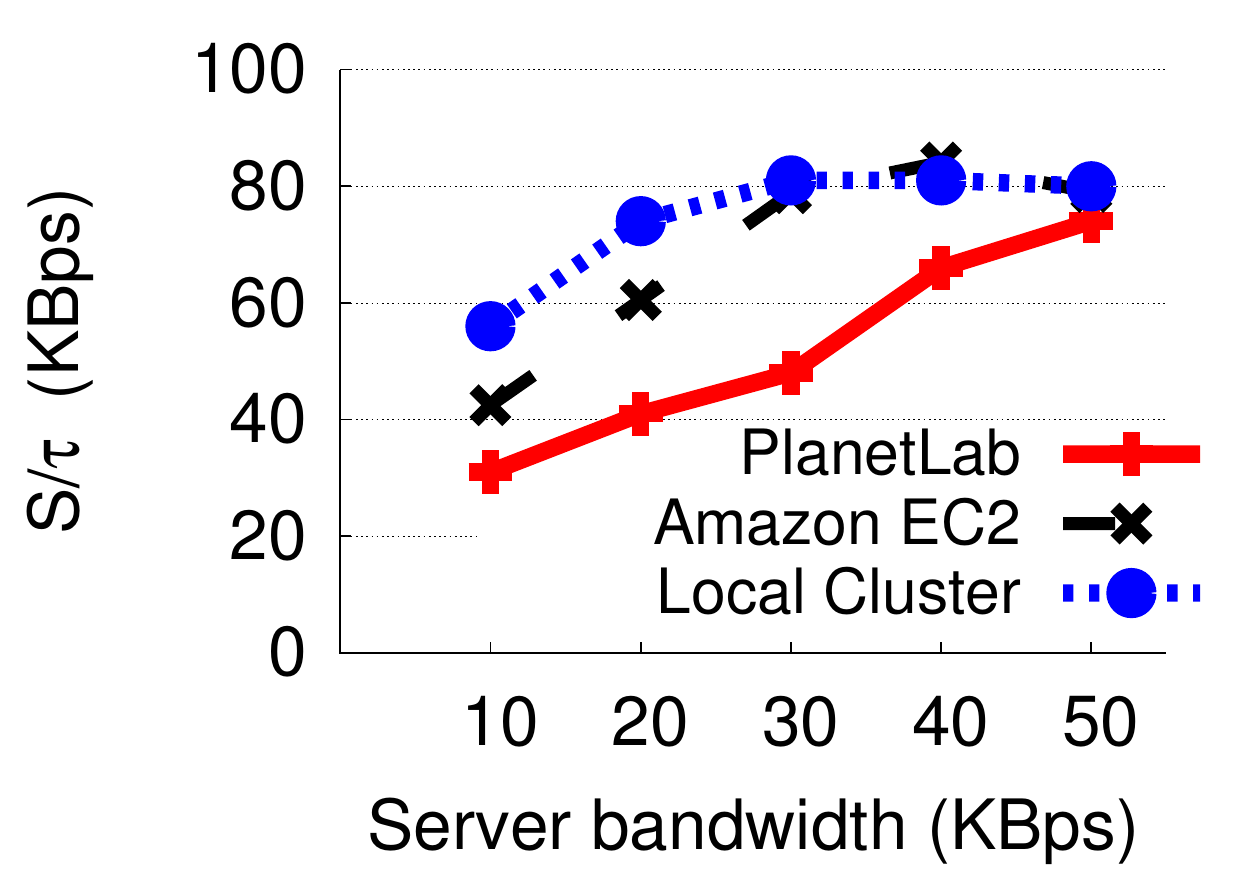}
\caption{Swarm performance is better on Amazon EC2 and on a local cluster compared to Planetlab.}
\label{fig:testbed}
\end{minipage}
\end{figure}
\vspace{-0.1in}


We address file size diversity using an interpolation approach similar to the one used for arrival rates and server bandwidth. A separate cheat sheet is stored for a small number of file sizes spanning the regime of interest, e.g., 10 file sizes in geometric progression from 1MB to 10GB. The swarm performance for file sizes in between is estimated via interpolation. 

\eat{
The cheat sheet as described above can only be used to estimate swarm performance for a given file size and upload capacity distribution. In practice, the peer upload capacity distribution at a managed swarming site is unlikely to change significantly over short time scales. Although the upload capacity distribution of any sample of peers currently participating in a swarm may differ from the overall distribution, the overall distribution depends on technology trends and the population of users who visit the site, which is likely to remain stable over the course of several months. To address file size diversity, an interpolation approach similar to the one used for arrival rates and server bandwidth is employed. A separate cheat sheet is stored for a small number of file sizes spanning the regime of interest, e.g., 10 file sizes in geometric progression from 1MB to 10GB, and the swarm performance for file sizes in between is estimated via interpolation. 
}

At the onset of this work, we expected that a larger file size could be treated as equivalent to a larger arrival rate, i.e., $f(x, \mu, \lambda, kS)$ could be approximated as $f(x, \mu, k\lambda, S)$, thereby obviating the need to maintain separate cheat sheets for different file sizes. Our intuition was that $\lambda S$ (bits/sec) represents the aggregate demand arriving into the system, so the response curve should not change significantly if the demand remains unchanged. Unfortunately, this turns out not to be the case as shown by the experiment in Figure \ref{fig:size}. The figure plots the swarm performance as a function of the server bandwidth, and the different lines increase (decrease) $S$ ($\lambda$) by the same factor, i.e., $\lambda S$ is the same for all points in the graph. The lines clearly show a slight uptrend suggesting that larger file sizes boost swarm performance more than larger arrival rates or, equivalently, a swarm distributing a larger file performs better than a swarm distributing a smaller file even though both have the same aggregate demand, client upload capacities, and server bandwidth.

\paragraph{Upload capacity distribution} There are two kinds of variations that occur in peer upload capacities. First, the upload capacity distribution of any sample of peers currently participating in a swarm may differ from the overall distribution. Our model implicitly accounts for this statistical variation  because peer upload capacities during measurements are chosen by randomly sampling the distribution. Second,  the overall upload capacity distribution of peers visiting the site can change. However, we expect that upload capacity distribution is unlikely to change at short time scales, as it depends on technology trends and the population of users who visit the site, which is likely to remain stable over the course of several months.

%
%

The changes in the upload capacity distribution at time scales of several months can be addressed by updating the cheat sheet with new measurements. In additional experiments (included in our tech report \cite{techreport}  due to lack of space), we find that significant changes in the mean or even the variance of upload capacity distribution indeed necessitate a new set of measurements. For example, for the same mean upload capacity, we find that increasing the variance of upload capacities reduces the swarm performance.

\subsubsection{Effect of measurement testbed}
\label{sec:network}

The measurement-based model requires network conditions to remain relatively similar to the environment in which the model's measurements were obtained.  We repeated the experiment shown in Figure \ref{fig:basic} on two other testbeds - Amazon EC2 \cite{amazonec2}, and a local cluster. For the EC2 experiment, we select equal number of machines from five geographic locations to differentiate the EC2 testbed from the local cluster which has microsecond round trip latencies.  In Figure \ref{fig:testbed}, we compared the swarm performance on the three testbeds for a peer arrival rate of $\lambda$ = 1/5/s. Swarm performance on EC2 is up to 30 KBps higher than on PlanetLab. Experiments on the local cluster show even better swarm performance than on EC2. 


Swarm performance differs on the three testbeds as their \emph{effective} upload capacities are different. The round-trip times in the local cluster are much smaller than in PlanetLab which reflects in the form of higher effective upload capacities and better performance. EC2 only has a small extent of geographic diversity (five different locations), so neighbor relationships between peers in the same data center tend to dominate (a clustering effect that has also been alluded to by prior work \cite{legout:07}). This clustering effect again results in the form of EC2 nodes having higher effective upload capacities.





\subsubsection{Summary and limitations}
\label{sec:summary}

Although the measurement-based model can capture the dependence on four key swarm parameters, it still has several limitations.  The most critical limitation is the extensive measurement needed to build a cheat sheet.  For a single file size, our measurement take a few hundred hours on PlanetLab. A content distributor maintaining a few such tables for common file sizes may require a few thousand hours of measurements or even more.

The second limitation is the difficulty in estimating two of the model parameters -- upload capacity distribution and peer arrival rates. Upload capacity distribution is difficult to estimate for several reasons -- peers may download files from multiple swarms simultaneously or otherwise limit their upload capacity, and network conditions can significantly change the effective upload capacity  as shown in Section \ref{sec:network}.  Estimating peer arrival rates is challenging primarily because users may abort the download before completion and return later to resume a download  as shown in prior work \cite{p2pchurn,youtubeUser}. Therefore, the model also needs to account for peer arrivals and departures in the middle of a download.  In combination, the difficulty of estimating all the model parameters can make the  measurement-based model ineffective in practice.

\section{\cs\ system}
\label{sec:swarmserver}

In this section, we present an implemented prototype of our system, \cs,   to compare different controller strategies.    We begin with a brief description of our implementation and the content distribution objectives that we use for our comparison.  Then, we discuss the design of model-based,  dynamic, and static controllers implemented in \cs.



\emph{Implementation:}
\label{sec:objectives}
\cs\ system is implemented in Python and consists of nearly 5000 lines of code. The system does not require any modification to the BitTorrent protocol for either the peers or the tracker. Our implementation uses  the instrumented BitTorrent client developed by Legout et al. \cite{legoutrf}, which we modified to enable us to change the maximum upload bandwidth of the client without restarting it.

\emph{Content distribution objectives:} We compare controller strategies on three content distribution objectives.


\begin{itemize}
\item
{\bf\maxavg}: Minimize the average download time across all peers in all swarms for a given total server bandwidth.
\item
{\bf\maxmin}: Minimize the maximum value of the average download time across swarms for a given total server bandwidth.
\item
{\bf\mincost}: Minimize the total server bandwidth while achieving a set of specified target download times for each swarm. 
\end{itemize}
\vspace{-0.1in}

\subsection{Model-based controller}
\label{sec:model-controller}

The model-based controller -- \sscs\ -- allocates server bandwidth by solving an optimization problem using the measurement-based model developed in Section \ref{sec:model-based}. Next, we describe the optimization formulations used by \sscs\ to calculate bandwidth allocation for each of the objectives introduced above. We assume that there are a total of $k$ swarms and the average upload capacity, arrival rate, and file size of the $i$'th swarm $1\leq i \leq k$ are given by $\lambda_i, \mu_i, S_i$ respectively. The goal is to determine server bandwidth allocations $\{x_i\}_{1\leq i\leq k}$ so as to optimize the desired objective.





\noindent\textbf{Optimization formulation for \maxavg:} 
\begin{eqnarray}
\min\sum_{1\leq i\leq k}{\lambda_i}{\tau_i}
\label{eq:maxrate}
\end{eqnarray}
\vspace{-0.1in}
subject to 
\begin{eqnarray}
\tau_i &= & S_i/f(x_i,{\mu}_i, \lambda_i, S_i),  \ \ \ \ 1\leq i \leq k
\label{eq:model_constraint}\\
\sum_{1\leq i\leq k}x_i &\leq& X
\label{eq:cap_constraint}
\end{eqnarray}

The first constraint (\ref{eq:model_constraint}) above simply rephrases Equation (\ref{eq:model}) relating the average download time $\tau$ to the server bandwidth $x$ and other swarm parameters. The second constraint above limits the total bandwidth the server can allocate to all swarms.

\sscs\ uses its measured knowledge of $f(.)$ to solve this optimization problem. If $f(.)$ is known to be smooth and concave in $x$, \maxavg\ can be solved using a greedy gradient-ascent strategy that computes a unique, optimal solution as follows: (1) Start with $x_1=x_2=\cdots=x_k=\Delta$ for a small $\Delta$; (2) Allocate the next $\Delta$ units of capacity (divided equally) to the swarm(s) with the largest value(s) of the gradient $\lambda_i f^{'}(x_i,{\mu}_i,\lambda_i, S_i)$; (3) If not all $X$ units of capacity have been allocated, goto (2). Else terminate. 

If $f(.)$ is piecewise linear and concave, the above strategy still works, but the resulting solution may not be unique.   In order to compute a unique optimal solution, \sscs\ cleans the measured $f(.)$ by fitting smooth and concave polynomial curves for each line in Figure \ref{fig:basic}. We assume that this data cleaning has been already performed while describing the solutions to the next two objectives as well.


\noindent\textbf{Optimization formulation for \maxmin:} 
\begin{eqnarray}
\min(\underset{1\leq i\leq k}{\max}(\tau_i))
\label{eq:maxmin}
\end{eqnarray}
subject to the same constraints as (\ref{eq:model_constraint}) and (\ref{eq:cap_constraint}) above.

If $f(.)$ monotonically increases with $x$, \maxmin\ can be solved optimally using a simple greedy heuristic. For a target rate $y$, let  $x = f^{-1}(y, \mu, \lambda, S)$ denote the server bandwidth $x$ required to achieve an average download time of $S/y$. The heuristic is as follows: (1) Initialize target rate $y=\Delta$ for a small $\Delta$; (2) Set $x_i=f^{-1}(y,{\mu_i},\lambda_i,S_i)$, $1\leq i\leq k$; (3) If bandwidth allocation required to achieve the target is feasible, i.e.,  ($\sum_i{x_i}< X$),  increment target rate $y$ to $y+\Delta$ and goto (2). Else, terminate.



\noindent\textbf{Optimization formulation for \mincost:} 
\begin{eqnarray}
\min(x_1+\cdots+x_k)
\label{eq:mincost}
\end{eqnarray}
\vspace{-0.1in}
subject to
\begin{eqnarray}
\tau_i = S/f(x_i,{\mu}_i, \lambda_i, S_i),  \ \ \ \ 1\leq i \leq k
\label{eq:targets}
\end{eqnarray}

If $f(.)$ is invertible, then \mincost\ can be solved by setting $x_i = f^{-1}(S/\tau_i, \mu, \lambda, S)$.

\subsection{Dynamic controller}
\label{sec:dynamic}

We implement three dynamic controllers: \ssaiad,  \ssmm, and \ssaf.

%

\ssaiad\  optimizes the \mincost\ objective and  works as follows. Suppose the target average download time of the swarm is $\tau$ and the file size is $S$. AIAD initializes the server bandwidth $x$ to $S/\tau$.  Once every epoch, it  measures the average download rate, $y$, of peers in the swarm. If  $S/\tau > y$,  it increases the server bandwidth $x$ by $\Delta$. Otherwise it decreases $x$ by $\Delta$, except in the case that the decrement would cause $x$ to dip below a minimum bandwidth threshold. Our implementation sets the epoch length to 200 s $\Delta$  to 10 KBps, and the minimum bandwidth threshold to 5 KBps.

\ssmm\  optimizes the \maxmin\ objective.   At the start, \ssmm\ assigns equal bandwidth to all swarms.  Once every epoch, \ssmm\ measures the average download rate of all swarms. The server bandwidth is increased by a small, fixed $\Delta$ for swarms whose download rate is lower than the median of average download rates. Similarly, \ssmm\ reduces the server bandwidth by $\Delta$ for each swarm with average download rate higher than the median value. Similar to \ssaiad, \ssmm\ never reduces the server bandwidth allocated to a swarm below a minimum threshold. Epoch length, $\Delta$, and the minimum bandwidth are the same as in \ssaiad.


\ssaf\ optimizes the \maxavg\ objective and is based on the algorithm in the AntFarm paper  \cite{antfarm}.  At the start, \ssaf\ allocates a small initial bandwidth to every swarm, and then assigns the server bandwidth to swarms in small increments until all server bandwidth is used up.

In steady state, \ssaf\ computes  the bandwidth allocation using \emph{response curves} for each swarm that predict the swarm performance as a function of server bandwidth. AntFarm measures download rates of peers periodically to obtain a set of  sample data points of the form (server bandwidth, swarm performance). The response curve for a swarm is computed by fitting a concave, piecewise-linear curve to this set of data points. Given the response curves for all swarms, their bandwidth allocation is determined using a gradient-ascent algorithm  similar to that used by \sscs's  to optimize the \maxavg\ objective. We refer the reader to our tech report \cite{techreport} for a detailed description of our implementation.

\subsection{Static controller}

We implement three static controllers. (1) {\em \btcap} sets an upload limit at the server for a set of swarms but does not set a per-swarm limit. The server bandwidth to each swarm by the server is determined by the number of peers connected to the server. (2) {\em \eqsplit} splits the available server bandwidth equally among all swarms. (3) {\em \propsplit} splits total server bandwidth proportional to the peer arrival rate for each swarm.  

\section{Evaluation}
\label{sec:eval}

\begin{figure*}[ht]
\vspace{0.05in}
\begin{minipage}[b]{0.3\linewidth}
\centering
\includegraphics[scale=0.44]{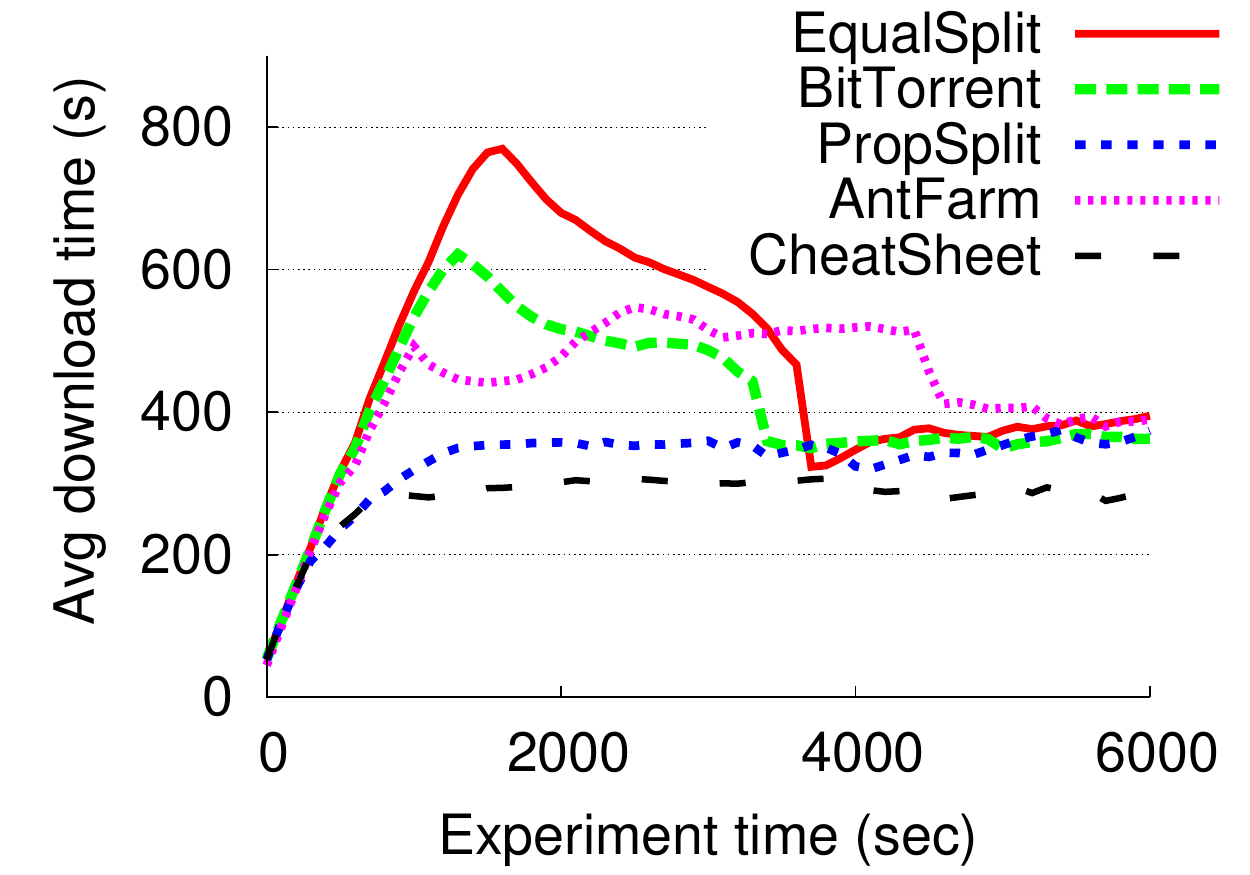}
\caption{For the \maxavg\ objective, both static  (e.g., Equal Spit) and dynamic controllers (e.g., AntFarm) incur a higher download time in the initial phase as well as in steady state.}
\label{fig:zipf-avg-time}
\end{minipage}
\hspace{0.5cm}
\begin{minipage}[b]{0.3\linewidth}
\centering
\includegraphics[scale=0.44]{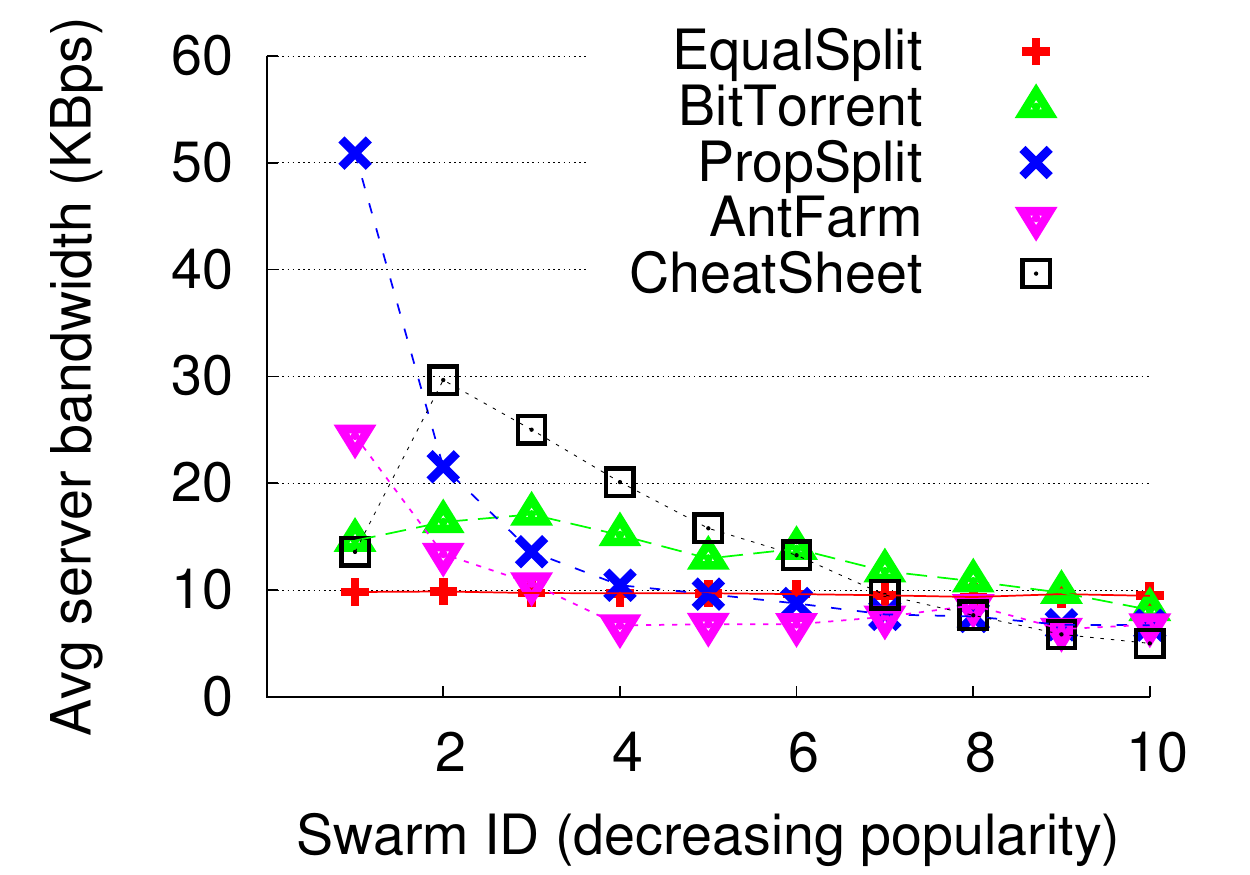}
\caption{Server bandwidth to swarms by controllers to minimize the average download time. Swarms are ordered left to right in decreasing order of popularity.}
\label{fig:zipf-avg-seeder-bw}
\end{minipage}
\hspace{0.5cm}
\begin{minipage}[b]{0.3\linewidth}
\centering
\includegraphics[scale=0.44]{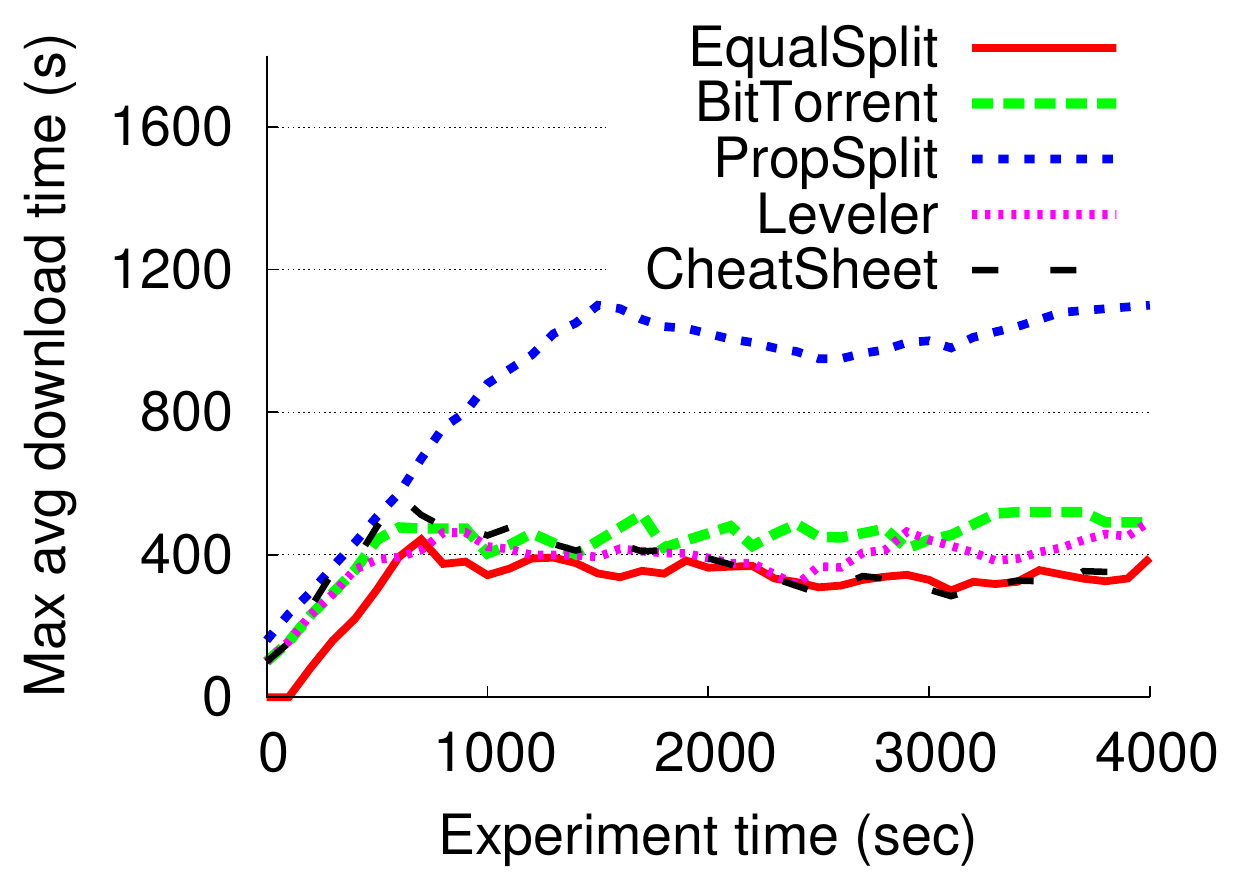}
\caption{For the \maxmin\ objective, \sscs\ and \eqsplit\ perform nearly the same, while \propsplit\ performs the worst.} 
\label{fig:zipf-maxmin-time}
\end{minipage}
\vspace{-0.25in}
\end{figure*}


Our comparison of controller strategies, presented in this section, answers two main questions: (1) Do dynamic and model-based controllers improve performance over static controllers? If yes, then by how much? (2) Which type of controller, dynamic or model-based, performs better for the objectives in Section \ref{sec:objectives}?  Our experiments show that model-based controller outperforms dynamic controllers on all three objectives we compared. Static controllers cannot equal a model-based controller either; they perform well on some workloads and objectives but fare poorly on others.

\textbf{Experimental setup:}  We performed our evaluation on 350 PlanetLab nodes using an instrumented BitTorrent client \cite{legoutrf}.  Upload and download capacities of peers and peer inter-arrival times follow the same distributions as in our measurements.

Due to limited upload capacities and daily data transfer limit on PlanetLab, we focus on experiments with small file sizes. File download time will increase in proportion to the file size as we cannot increase upload capacity significantly. Thus, an experiment with a 1 GB  file could take 100$\times$ longer to finish  than an experiment with a 10 MB file,  if the same number of file downloads occur in both experiments (assuming both swarms have same aggregate demand $\lambda S$).

\subsection{Average download time}


First, we compare controllers on the \maxavg\ objective.  We select a workload consisting of 20 swarms whose mean arrival rates are chosen according to a Zipf distribution with parameter 1.5.  The mean arrival rate of the most popular swarm and the least popular swarm is 0.5/s and 0.0055/s respectively. 
Each swarm distributes a file of size 10 MB. The total server bandwidth is set to 200 KBps.


%
%

Figure \ref{fig:zipf-avg-time} shows how the average download time changes over time for the different compared schemes.  The average is computed using the download time of peers that completed their download within the previous 2000 sec interval as well as the resident time, i.e., the time since arrival, for peers whose downloads are under progress.

There are two main observations from the experiment in Figure \ref{fig:zipf-avg-time}. First, in the initial phase, \eqsplit, \btcap\ and \ssaf\ incur much higher average download times than \propsplit\ and \sscs,  and their average download times take considerably longer to stabilize. Second,  even after all controllers have reached steady state, \sscs\  achieves a download time that continues to be lower (by at least 25\%) compared to all other schemes (that perform roughly similarly in steady state in this experiment).




The explanation for these observations is as follows.  \eqsplit, BitTorrent and AntFarm have a very  high download time  at the start of the experiment because they assign a small server bandwidth to large and small swarms alike. If the initial server bandwidth is small, a huge number of peers build up in highly popular swarms, which is reflected in the corresponding download time curves that rise rapidly. For example, the download times in \eqsplit\ increase rapidly until about 1500 sec as no peers have departed until then. At this point, the download time drops sharply as a result of a horde of peer departures that occur when the last block in a swarm has been uploaded by the server. In contrast, both \sscs\ and \propsplit\ assign higher bandwidth to popular swarms from the start, so peer departures start much quicker in popular swarms considerably reducing their average download times. We note here that \sscs\ is implemented so as to begin with an allocation identical to \propsplit\ until it has a stable estimate of peer arrival rates, at which point it switches to the model-based optimal allocation. 


\eqsplit, \btcap\ and \ssaf\ take considerably longer to reach steady state because the number of peers in highly popular swarms goes through multiple rounds of ramp-ups followed by bulk departures before stabilizing.  In this experiment, \ssaf\  takes the longest time to converge to a steady state because after assigning 5 KBps to each swarm at the beginning, it allocates remaining bandwidth in small chunks of 5 KBps once every 200 sec. \ssaf\ requires many such 200 sec epochs in order to build a stable response curve for all swarms, resulting in higher download times during this convergence phase. We have observed (not shown for brevity) that reducing the epoch length does not help and sometimes hurts performance as it increases the measurement error in learned response curves.


Why does \sscs\ outperform other schemes even in steady state? Figure \ref{fig:zipf-avg-seeder-bw} shows the steady-state allocations of server bandwidth achieved by different schemes that explain this observation. Swarms are ordered from left to right in decreasing order of popularity. We show only the top 10 most popular swarms for  clarity of presentation. \sscs\ uses the model to predict that the most popular swarm is mostly self-sustaining and therefore needs only a small bandwidth to achieve healthy download times. Compared to other controllers, \sscs\ assigns higher bandwidth values to the next four popular swarms that belong to a regime where a small amount of server bandwidth disproportionately improves performance, which considerably reduces the average download time. \propsplit\ and \ssaf\ by design assign the most bandwidth to the most popular swarm, but the extra server bandwidth hardly benefits that swarm. \btcap\ is biased more towards the popular swarms (as it receives more peer connections from these swarms compared to singleton swarms), but its allocation is nevertheless sub-optimal. \eqsplit\ clearly makes a sub-optimal decision by allocating equal bandwidth to all swarms in the light of the above reasons.

\subsection{Min-max average download time}

Next, we compare controllers on the \maxmin\ objective, i.e., minimizing the average download time of the swarm that has the worst average download time. We evaluate on the Zipf workload in the previous subsection and set the total server bandwidth to 500 KBps in these experiments. 



Figure \ref{fig:zipf-maxmin-time} shows the average download time of the swarm with the maximum average download time (referred to as MAD time in this discussion). We observe that, even though the workload is the same, the relative performance of controllers is different compared to the experiment in the previous subsection. The MAD time achieved by \propsplit\ is twice as worse as other controllers that have relatively smaller differences between them. Both \eqsplit\ and \sscs\ achieve the lowest MAD time. BitTorrent incurs a higher MAD time in comparison to \eqsplit. The performance of \ssmm\ varies with time because it changes  the server bandwidth to each swarm periodically and struggles to converge to a steady bandwidth allocation as it shuffles bandwidth across 20 swarms. The reason (not visible in the figure) is that different swarms take different times to manifest the effect of the most recent change. \ssmm\ sometimes ``panics'' and allocates more bandwidth to the currently worst swarms too quickly and at other times is too slow to move bandwidth away from swarms that could do without it. The fluctuating performance of \ssmm\ reveals that it is nontrivial to design a robust dynamic controller.

Unpopular swarms, i.e., swarms with a small peer arrival rate, impact the MAD time significantly in this experiment. Unpopular swarms require higher bandwidth than popular swarms to achieve the same download time (Figure \ref{fig:basic}).  Due to the Zipf popularity distribution, a majority of swarms for this workload are unpopular. \propsplit\ incurs the highest MAD times because it assigns the least bandwidth to the most unpopular swarm, which significantly increases the download time of that swarm. \eqsplit, unlike \propsplit, assigns equal bandwidth to all the swarms and hence has a much smaller MAD time. \sscs\  performs the same as \eqsplit\ because  the unpopular swarms in the workload have nearly the same performance in both cases. Due to a large number of unpopular swarms,  \sscs\ only assigns 5 KBps more bandwidth to each unpopular swarm than \eqsplit, which does not sufficiently impact the MAD times.

Does EqualSplit achieve the least MAD time for all workloads? The answer is no. On a workload dominated by popular swarms, \eqsplit\ results in 50\% higher MAD time than \sscs\ (refer to tech report \cite{techreport} for this experiment). This experiment illustrates that performance of static controllers such as EqualSplit can vary depending on the workload.

\begin{figure}[t]
 \begin{center}
   \subfigure[$\lambda = 0.5/s$]{\label{fig:g1-time}\includegraphics[scale=0.34]{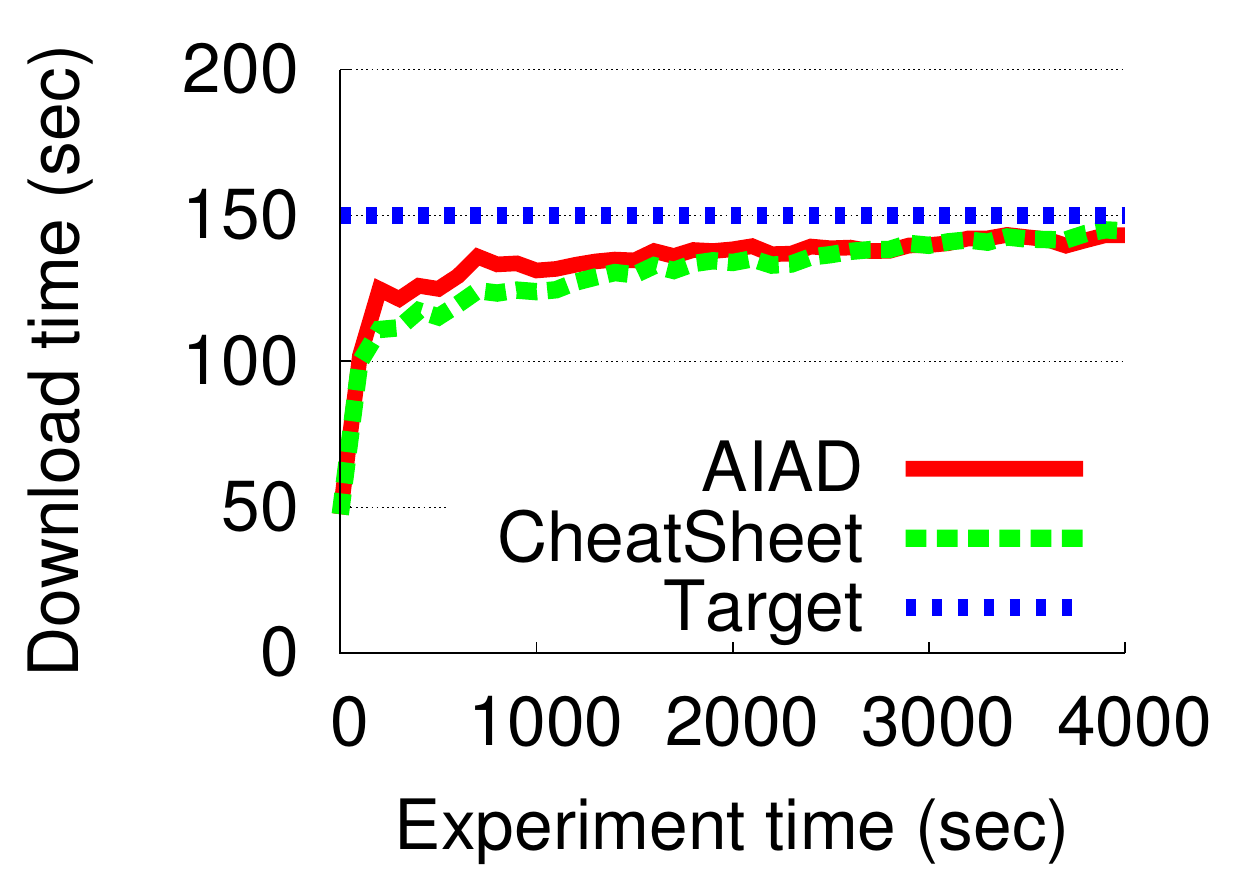}}
   \subfigure[$\lambda = 0.5/s$]{\label{fig:g1-uploads}\includegraphics[scale=0.34]{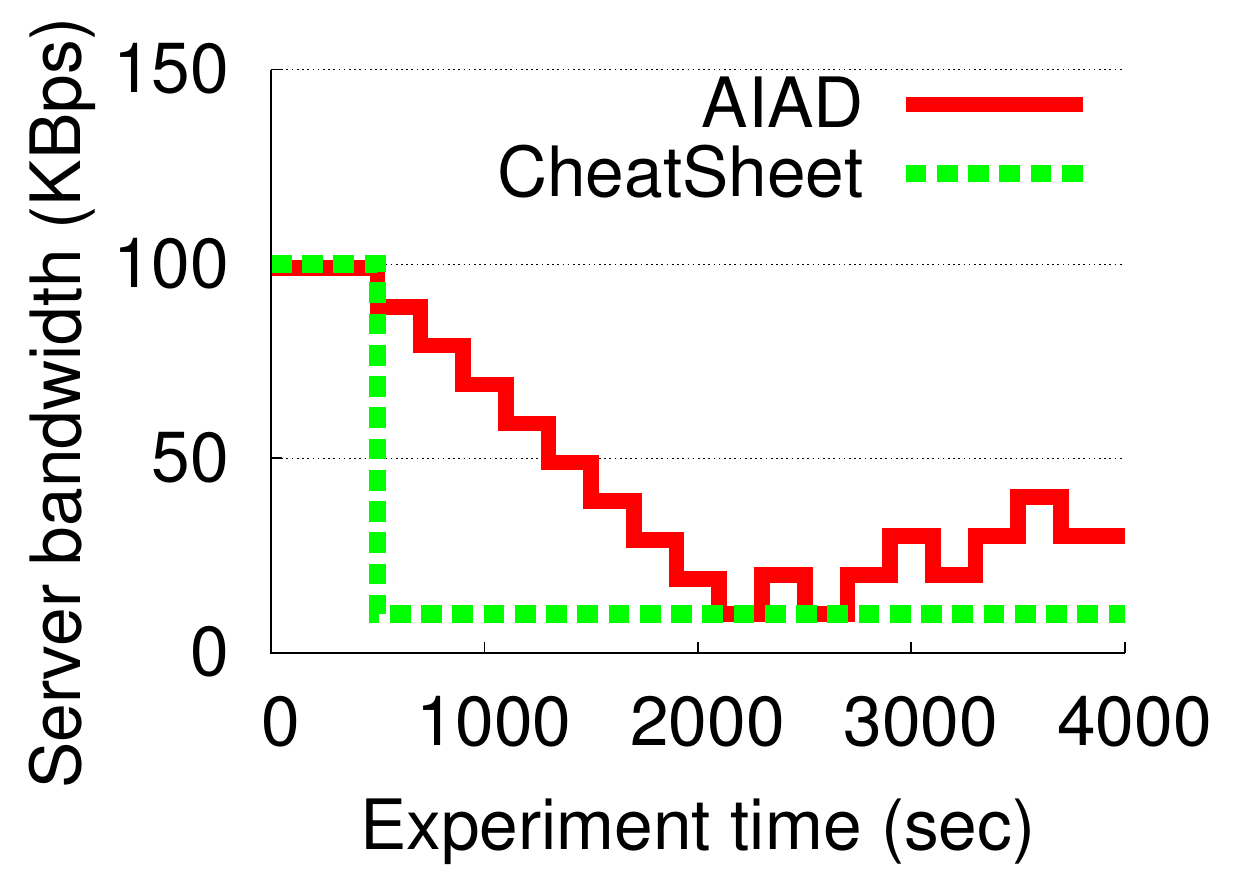}}
    \subfigure[$\lambda = 0.12/s$]{\label{fig:g2-time}\includegraphics[scale=0.34]{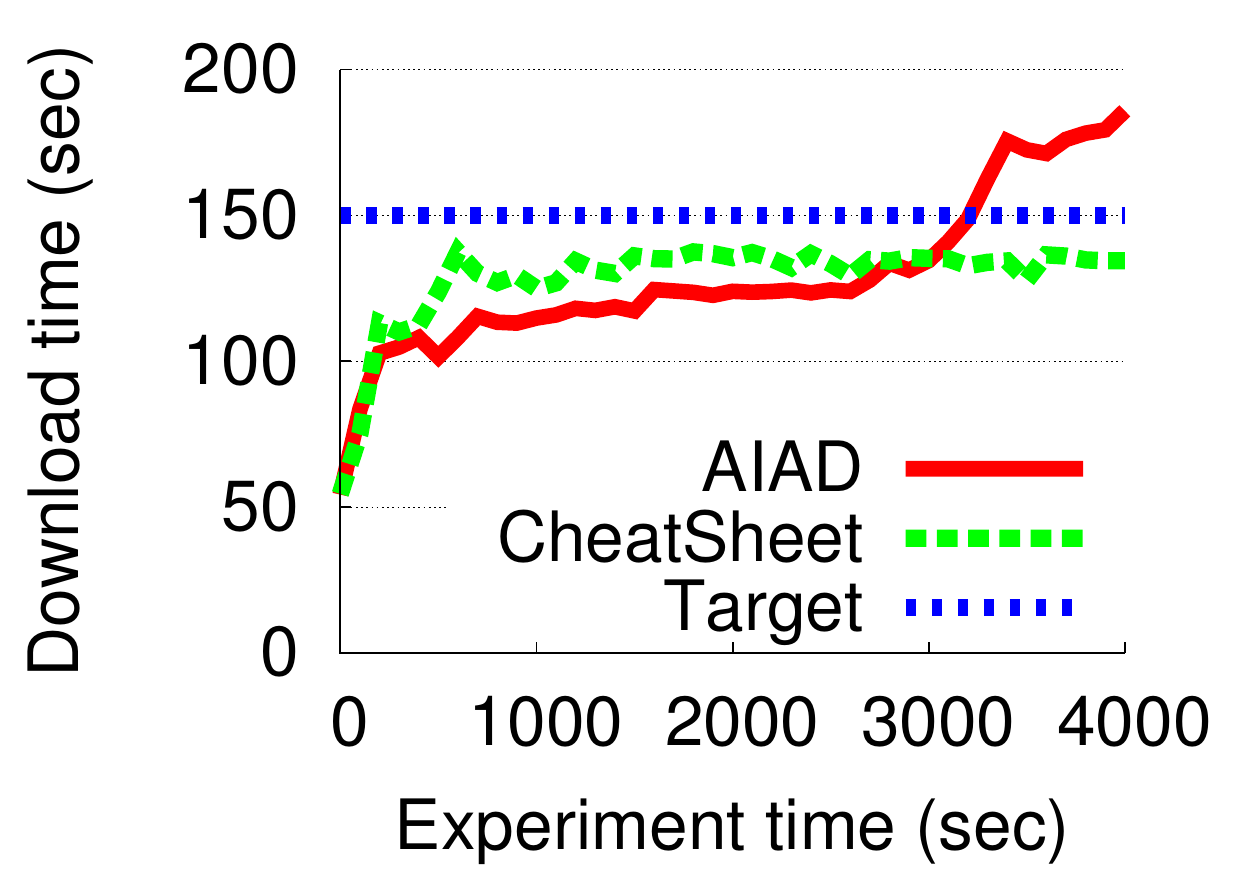}}
   \subfigure[$\lambda = 0.12/s$]{\label{fig:g2-uploads}\includegraphics[scale=0.34]{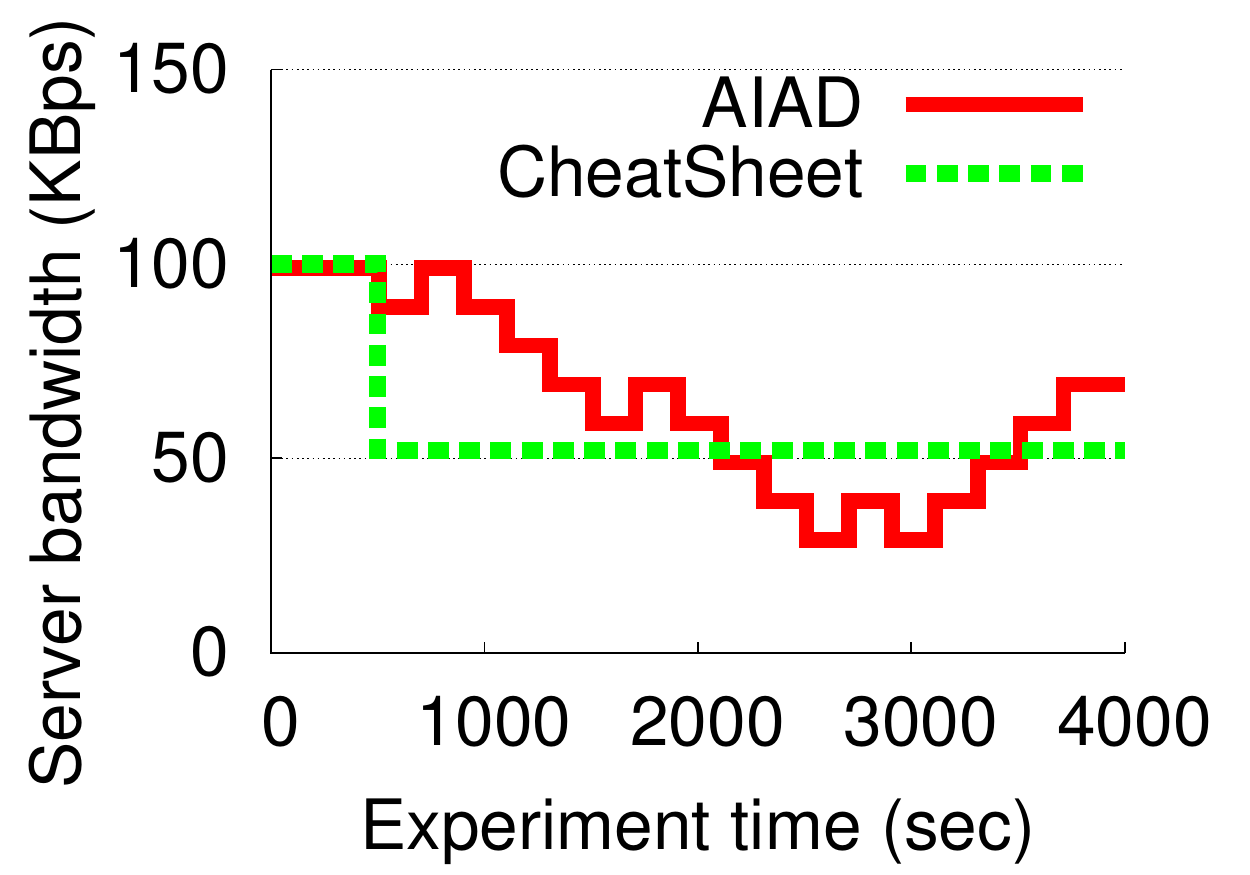}}
   \subfigure[$\lambda = 0.01/s$]{\label{fig:g3-time}\includegraphics[scale=0.34]{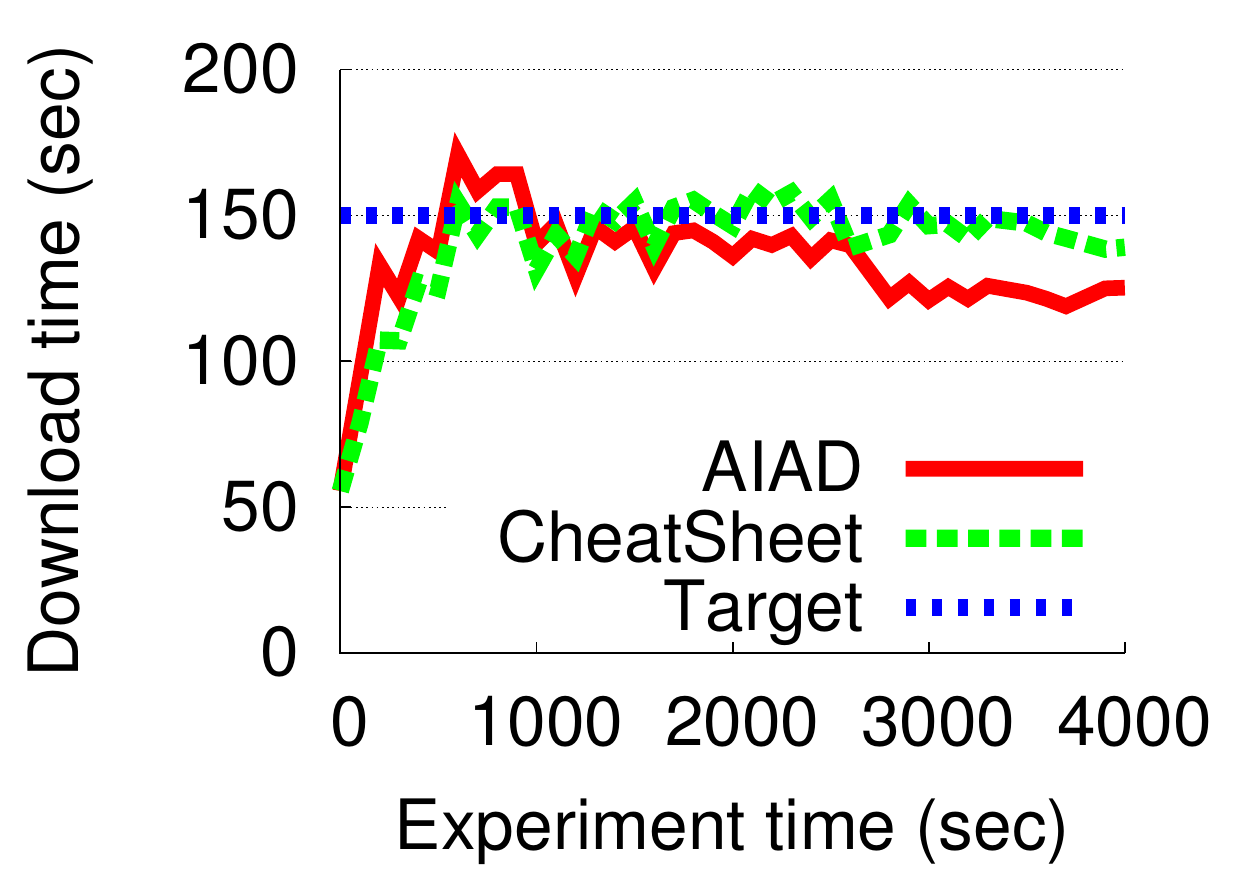}}
   \subfigure[$\lambda = 0.01/s$]{\label{fig:g3-uploads}\includegraphics[scale=0.34]{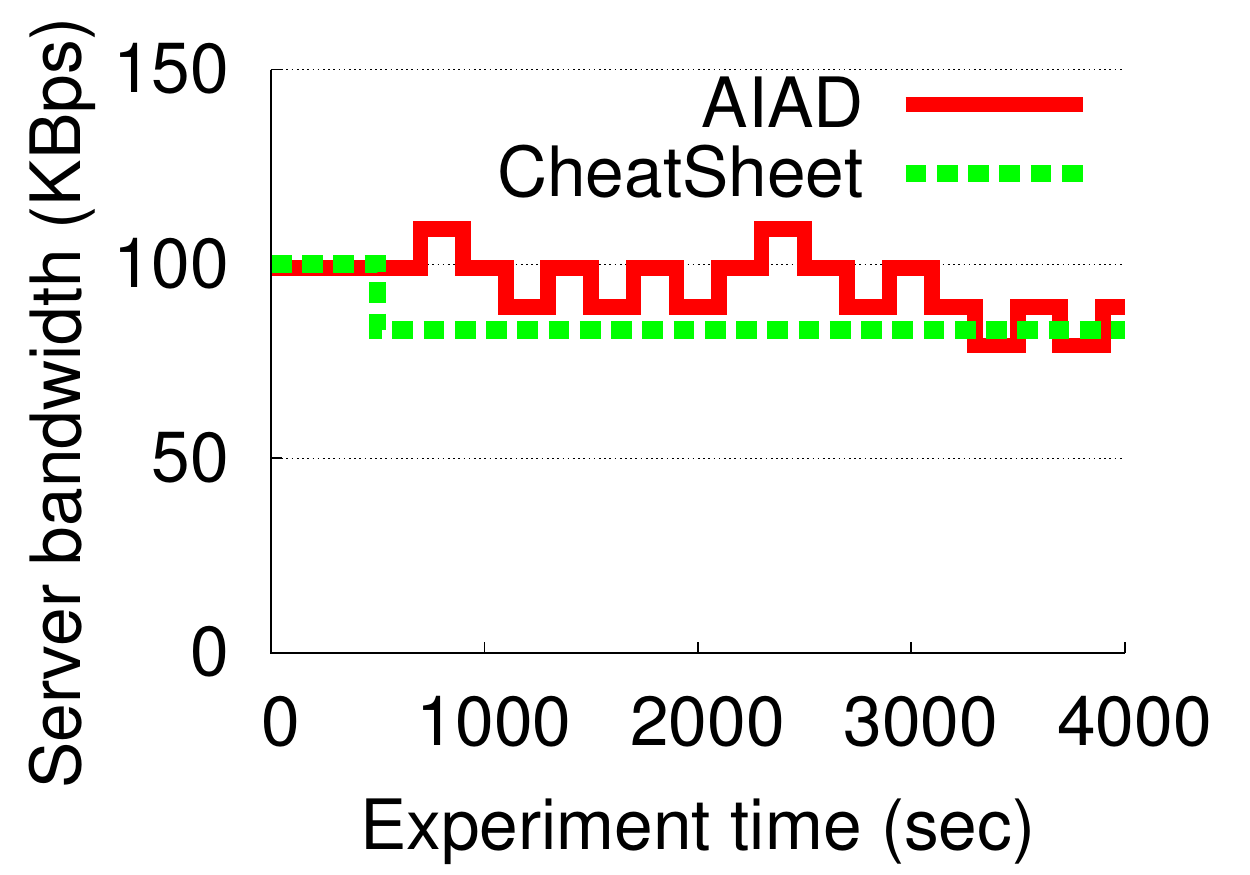}}
    \end{center}
     \vspace{-0.15in}
 \caption{Server bandwidth set by controllers.  \ssaiad\ fails to meet the target for $\lambda=0.12/s$ as it drastically reduces server bandwidth near 3000 s.}
 \label{fig:target}
 \vspace{-0.2in}
\end{figure}

\subsection{Target download time}

Next, we compare \sscs\ and \ssaiad\ against the \mincost\ objective. We do not compare against the simplistic static schemes as they are designed to always use all available capacity (and can therefore be made to appear arbitrarily worse by choosing a sufficiently low target download time in an experiment).  Our workload for this experiment consisted of six swarms with peer arrival rates of  0.5/s, 0.14/s, 0.12/s, 0.1/s, 0.08/s, and 0.01/s. All swarms distributed a file of size 10 MB. The target download time for all the swarms is set to 150 sec. We only present detailed results for arrival rates 0.5/s, 0.12/s, and 0.01/s here.  Results for other arrival rates are qualitatively consistent and are omitted due to lack of space.

Figure \ref{fig:target} shows the average download time achieved by each strategy over the duration of the experiment (figures on the left column) and shows the corresponding server capacity set by the controllers over the same duration (figures on the right column). The actual bandwidth consumed at the server is very close to the configured capacity shown in the figure.


\begin{wrapfigure}{r}{0.24\textwidth}
  \begin{center}
    \vspace{-0.2in}
\includegraphics[scale=0.33]{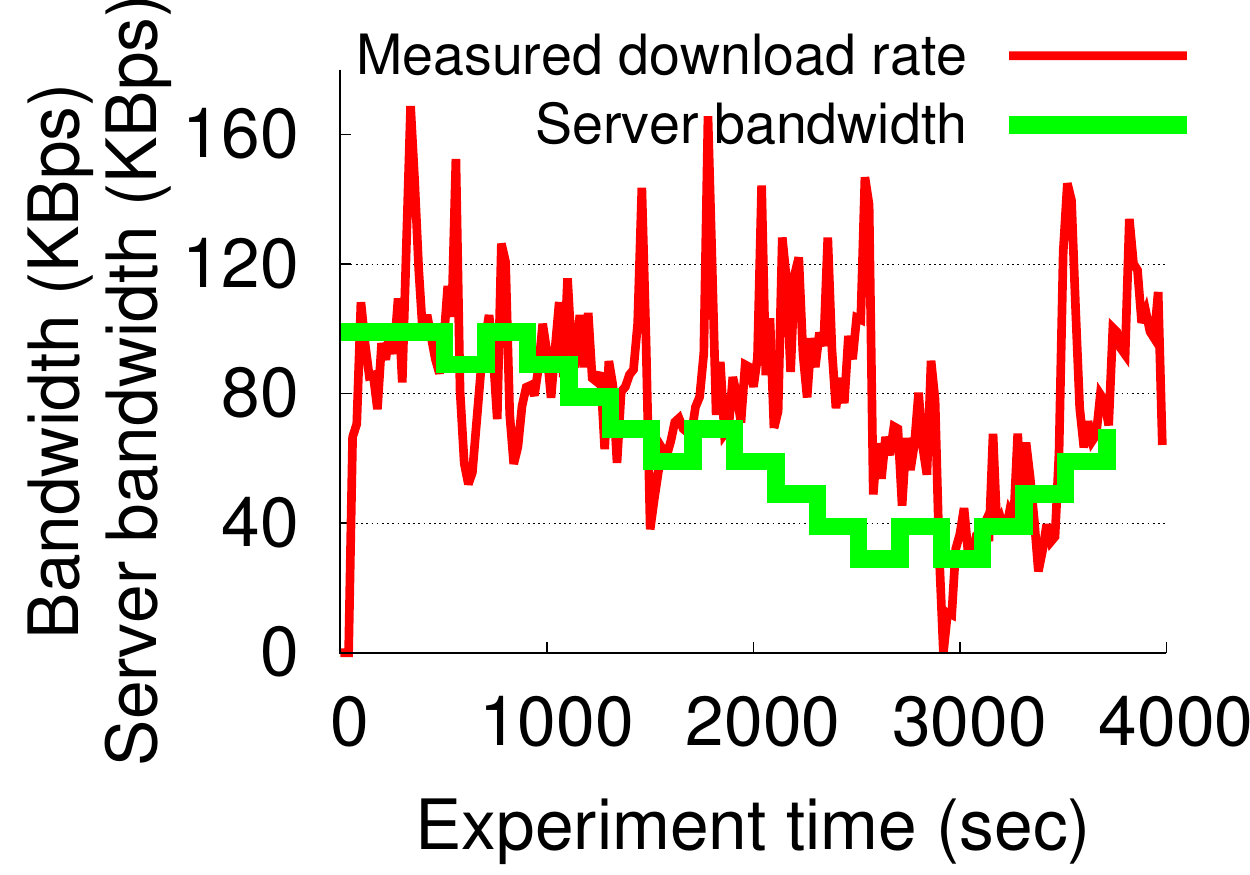}
  \end{center}
  \vspace{-0.17in}
  \caption{Server bandwidth set by \ssaiad\ in response to measured download rates. $\lambda$ = 0.12/s.}\label{fig:g2-downup}
 \vspace{-0.1in}
\end{wrapfigure} 
\sscs\ meets the target download time well in all cases, but \ssaiad\ sometimes significantly exceeds the target download time as in the later part of Figure \ref{fig:target}(c). This is because \ssaiad\ is not always able to accurately estimate the relation between server bandwidth and the download time. To illustrate this point, Figure \ref{fig:g2-downup} shows the measured download rate of the swarm and  the server bandwidth limit set by AIAD during this experiment.  At t = 2200 s, the measured download rate of swarm is above the corresponding target download time (10 MB / 150 s = 67 KBps). Hence it decreases the server bandwidth to 40 KBps at t = 2200 s and then to 30 KBps at t = 2400 s. This causes the measured download rate to drop sharply which is reflected in the increased download time of peers in Figure \ref{fig:g2-time}. The download time curve shows an increase somewhat later as it is calculated as an average over a window of 2000 s.

We also experimented by changing the interval after which \ssaiad\ updates bandwidth to 300 sec,  but it continues to fluctuate above the target download time.  Of course, if the bandwidth update interval is increased to a sufficiently high value and the bandwidth increments/decrements made small, the AIAD controller will converge to the target download rate. However, it will take longer to converge and will be less responsive if peer arrival rates change.

\sscs\ consumes much less bandwidth compared to \ssaiad\  for $\lambda$ = 0.5/s especially in the first 2000 seconds of the experiment. While \ssaiad\ takes several cycles of measurement and perturbations to reach the bandwidth allocation, \sscs\ directly jumps to the minimal required bandwidth using its model.

\subsection{Summary and discussion}

In summary, our evaluation shows that bandwidth allocation done by static controllers is hit-or-miss. A static controller that works well for one objective and workload combination may perform poorly for others. This is intuitively unsurprising and is also consistent with the findings in prior work analyzing a specific optimization objective \cite{antfarm}. For a fixed performance objective however, the simplicity of static controllers may outweigh their sub-optimality (e.g., \eqsplit\ for the \maxmin\ metric or \propsplit\ for the  \maxavg\ metric).

Our evaluation also shows that designing a dynamic controller for scenarios involving peer arrivals  and departures is nontrivial. Although  dynamic controllers are generally superior to any given simplistic static scheme when evaluated over a range of objectives and workloads, we find that they are far from optimal. Indeed, in some scenarios, simple schemes like \eqsplit\ or \propsplit\ outperform dynamic control. The reason is that measuring the relationship between swarm performance and allocated bandwidth in an online manner is nontrivial. As a result of measurement errors, a dynamic control scheme is vulnerable to prolonged convergence delays or persistent fluctuations.

The experiments in this paper suggest that a model-based approach is feasible and promising. We find that when a model-based controller is given a cheat sheet based on prior measurements in the regime of interest, it consistently outperforms both static   and dynamic controllers for different  objectives and workloads.

Nevertheless, having gone through the experience of making a model-based controller work, our conclusion is that, in its current form, the complexity of the model-based approach outweighs its advantages. The extensive set of measurements required to build the model, reduce the viability of this approach. Further,  the challenges in estimating model parameters such as peer arrival rates and upload capacity distribution of peers  (see Section \ref{sec:summary}) can reduce the effectiveness of model-based approach.

\section{Related work}
\label{sec:related2}

Our primary contribution is a comparative analysis of different categories of bandwidth controllers for \cacd\ systems and the design and implementation of a model-based control approach, that to our knowledge has not been attempted before. Our work builds upon a large body of prior work that can be grouped into dynamic controllers, models of swarm behavior, and incentive strategies.

\textbf{Dynamic controllers:} AntFarm  \cite{antfarm} and VFormation \cite{vformation} are closely related to ours. However, both these works adopt a dynamic controller approach. While AntFarm monitors average download rate of each peer, VFormation uses more detailed measurements by monitoring propagation of each block through the swarm. Our comparison of controller strategies does not include V-Formation because its implementation is proprietary and not available publicly. Dynamic controllers have also been studied for of live-streaming P2P systems \cite{p2ponline}. 


\textbf{Models of swarm behavior:} Qiu \cite{qiusrikant}, Fan \cite{fanchiu}, and Liao \cite{Liao} analytically model BitTorrent to derive expressions for average download time and other swarm metrics. But, their models  make assumptions that over-simplify swarm behavior, e.g., homogenous upload capacities \cite{qiusrikant}, fixed number of peers \cite{Liao}, and seeds contributing their full upload capacity \cite{qiusrikant,fanchiu}. To address these concerns, we model swarm performance based on actual swarm measurements.

\textbf{Incentive strategies:}
Several BitTorrent clients that improve BitTorrent's incentive strategies have been proposed, such as  BitTyrant \cite{bittyrant}, Levin et al.'s client \cite{levin08} and FairTorrent \cite{fairtorrent}.  Other swarming systems (incompatible with BitTorrent)  incentivize peers to contribute bandwidth  through  virtual currencies, e.g., Dandelion \cite{dandelion}, or tokens, e.g., AntFarm \cite{antfarm}. Our position is that a large majority of users use BitTorrent clients as-is or use unmodifiable closed-source clients, e.g.,  Akamai's NetSession \cite{akamai}, so incentive issues are less important.

\section{Conclusions}
\label{sec:concl}

In this paper, we performed a comparative evaluation of strategies to control server bandwidth  in \cacd\ systems. As part of this effort, we introduced a new approach referred to as model-based control and presented the design and implementation of a model-based controller, CheatSheet, that uses a concise model based on a priori offline measurement of swarm performance as a function of the server bandwidth and other swarm parameters. Our experiments show that simple static strategies are unreliable as they perform well on some workloads and objectives but fare poorly on others.  Dynamic control can also lead to a sub-optimal performance as it is prone to prolonged convergence delays and persistent fluctuations. In comparison, a model-based approach consistently outperforms both static and dynamic approaches provided it has access to detailed measurements in the regime of interest. Nevertheless, the broad applicability of a model-based approach may be limited in practice because of the overhead of developing and maintaining a comprehensive measurement-based model of swarm performance in each regime of interest.

\begin{small}
\bibliographystyle{IEEEtran}
\bibliography{IEEEabrv,New}


\end{small}

\end{document}